\documentclass[AMA,STIX1COL]{WileyNJD-v2}
\usepackage{booktabs}
\usepackage{multirow}
\usepackage{footnote}
\usepackage{float}
\usepackage{comment}
\usepackage{amsmath}
\usepackage{amsfonts}
\usepackage{titlesec}
\usepackage{bm}

\usepackage{graphicx,comment}
\usepackage{float}
\usepackage{url}
\usepackage{color}
\articletype{Research Article}%

\received{26 April 2016}
\revised{6 June 2016}
\accepted{6 June 2016}

\newcommand{\be}{\begin{eqnarray}}
\newcommand{\ee}{\end{eqnarray}}
\newcommand{\bee}{\begin{eqnarray*}}
\newcommand{\eee}{\end{eqnarray*}}
\newcommand{\bi}{\begin{enumerate}}
\newcommand{\ei}{\end{enumerate}}

\newcommand{\independent}{\perp\!\!\!\perp}

\usepackage[noabbrev,capitalize]{cleveref}
\crefformat{equation}{(#2#1#3)}

\newtheorem{prop}{Proposition}

\usepackage{xargs, xstring}
\newcommand{\E}{\mathbb{E}}

\allowdisplaybreaks

\begin{document}

\title{Propensity Score Weighting for Causal Subgroup Analysis}

\author[1]{Siyun Yang}
\author[2]{Elizabeth Lorenzi}
\author[3]{Georgia Papadogeorgou}
\author[4]{Daniel M. Wojdyla}
\author[5]{Fan Li}
\author[1,4]{Laine E. Thomas*}

\authormark{Yang \textsc{et al.}}

\address[1]{\orgdiv{Department of Biostatistics and Bioinformatics}, \orgname{Duke University School of Medicine}, \orgaddress{\state{North Carolina}, \country{USA}}}

\address[2]{\orgname{Berry Consultants}, \orgaddress{\state{Texas}, \country{USA}}}

\address[3]{\orgdiv{Department of Statistics}, \orgname{University of Florida}, \orgaddress{\state{Florida}, \country{USA}}}

\address[4]{\orgname{Duke Clinical Research Institute}, \orgaddress{\state{North Carolina}, \country{USA}}}
\address[5]{\orgdiv{Department of Statistical Science}, \orgname{Duke University}, \orgaddress{\state{North Carolina}, \country{USA}}}

\corres{
*Laine Thomas\\
  \email{laine.thomas@duke.edu}}

% SIYUN TO DO 
% Check SIM formatting requirements - should equations be referenced with \eqref or \ref? 
% SY: I used \eqref in the HTE sample size paper, which works fine.

% Clean up all references to appendixes - make sure the Appendix is referred to in the same language throughout the paper and with reference to the correct sections. 
% SY: this is fixed.

%\presentaddress{}

\abstract[Summary]{A common goal in comparative effectiveness research is to estimate treatment effects on pre-specified subpopulations of patients. Though widely used in medical research, causal inference methods for such subgroup analysis remain underdeveloped, particularly in observational studies. In this article, we develop a suite of analytical methods and visualization tools for causal subgroup analysis. First, we introduce the estimand of subgroup weighted average treatment effect and provide the corresponding propensity score weighting estimator. We show that balancing covariates within a subgroup bounds the bias of the estimator of subgroup causal effects. Second, we propose to use the overlap weighting method to achieve exact balance within subgroups. We further propose a method that combines overlap weighting and LASSO, to balance the bias-variance tradeoff in subgroup analysis. Finally, we design a new diagnostic graph---the Connect-S plot---for visualizing the subgroup covariate balance. Extensive simulation studies are presented to compare the proposed method with several existing methods. We apply the proposed methods to the Patient-centered Results for Uterine Fibroids (COMPARE-UF) registry data to evaluate alternative management options for uterine fibroids for relief of symptoms and quality of life.
}


\jnlcitation{\cname{%
  \author{Yang S.}, 
  \author{Lorenzi E.}, 
  \author{Papadogeorgou G}, 
  \author{Wojdyla D.}, 
  \author{Li F.} and 
  \author{Thomas L.E.}} (\cyear{2020}), 
  \ctitle{Propensity Score Weighting for Subgroup Analysis}, \cjournal{Statistics in Medicine}, \cvol{0000}.}

\maketitle

\section{Introduction}
\label{s:intro}

Comparative effectiveness research (CER) aims to estimate the causal effect of a treatment(s) in comparison to alternatives, unconfounded by differences between characteristics of subjects. CER has traditionally focused on the average treatment effect (ATE) for the overall population. However, different subpopulations of patients may respond to the same treatment differently, \citep{kent2007limitations,kent2010assessing} and in recent years the CER literature has increasingly shifted attention to heterogeneous treatment effects (HTE). \citep{hill2011bayesian,imai2013estimating, schnell2016bayesian, wager2018estimation, lee2018discovering} In particular, recent research employs machine learning methods to directly model the outcome function and consequently identify the subpopulations with significant HTEs \emph{post analysis}. Popular examples include the Bayesian additive regression trees (BART), \cite{chipman2010bart,hill2011bayesian} Causal Forest, \cite{wager2018estimation} and Causal boosting. \cite{powers2018}  In this article, we focus on a different type of HTE analysis, widely used in medical research: the 
causal {\it subgroup analysis} (SGA) which estimates treatment effects in \emph{pre-specified}---usually defined using pre-treatment covariates---subgroups of patients. There is an extensive literature on SGA methods in randomized controlled trials.  \cite{Assmann2000,Pocock2002,Wang2007, varadhan2014, Alosh2017} However, causal inference methods for SGA with observational data remain underdeveloped. \cite{Radice2012,dong2019subgroup,ben2020varying} 

In the context of ATE, covariate balance has been shown to be crucial to unbiased estimation of causal effects.\cite{imai2014covariate, zubizarreta2015stable} Propensity score methods\cite{rosenbaum1983central} are the most popular method for achieving covariate balance, but have seldom been discussed in SGA.\cite{Radice2012,dong2019subgroup} Compared to the aforementioned machine learning methods that directly model the outcomes, propensity score methods are design-based in the sense that they avoid modeling the outcome, and the quality of the analysis can be checked through balance diagnostics.\cite{rubin2008objective} In this paper we focus on the propensity score weighting approach.\cite{RobinsRotnitzky95, Robins2000, Hirano2001, Hirano2003, lif18, zhao2019covariate}  Dong et al. \cite{dong2019subgroup} shows that the true propensity score balances the covariates in expectation between treatment groups in both the overall population and any subgroup defined by covariates. However, the propensity scores are usually unknown in observational studies and must be first estimated from the study sample, leading to estimated propensity scores that rarely coincide with their true values. Moreover, good balance in the overall sample does not automatically translate in good subgroup balance. In fact, our own experience suggests that severe covariate imbalance in subgroups is common in real applications, which may consequently lead to bias in estimating the subgroup causal effects. 
Despite routinely reporting effects in pre-specified subgroups, medical studies rarely check subgroup balance, partially due to the lack of visualization tools.   
Indeed, we conducted a literature review of all propensity-score-based comparative effectiveness analyses
published in the \emph{Journal of American Medical Association (JAMA)} between January 1, 2017 and August 1, 2018. Of 16 relevant publications, half reported SGA (2-22 subgroups per paper) but {\it none} reported any metrics of balance within subgroups. 

The limited literature on propensity score methods in SGA suggests that the propensity score model should be iteratively updated to include covariate-subgroup interactions until subgroup balance is achieved. \citep{green14, wang18} But this procedure has not been implemented in practice, perhaps because it is cumbersome to manually check interactions. More importantly, it may amplify the classic bias-variance tradeoff: increasing complexity of the propensity score model may help to reduce bias but is also expected to increase variance. Therefore, an effective approach would automatically achieve covariate balance in subgroups while preserving precision. Machine learning methods offer a potential solution for estimating the propensity scores without pre-specifying necessary interactions. For example, generalized boosted models (GBM) have been advocated as a flexible, data-adaptive method,\citep{mccaffrey04} and random forest was superior to many other tree-based methods for propensity score estimation in extensive simulation studies.\citep{lee10} BART have been used to estimate the propensity score model and outperformed GBM on some metrics of balance. \citep{hill2011challenges}  However, it is unclear whether these methods achieve adequate balance and precision in causal SGA.  Moreover, when important subgroups are pre-specified, a more effective approach would incorporate prior knowledge about the subgroups.

In this article, we develop a suite of analytical and visualization tools for causal SGA. First, we introduce the estimand of subgroup weighted average treatment effect (S-WATE) and provide the corresponding propensity score weighting estimator (Section \ref{sec:estimand}). Second, we propose a method that combines LASSO \cite{tibshirani1996regression, belloni13} and overlap weighting, \cite{lif18,li2019addressing,thomas2020overlap} and balances the bias-variance tradeoff in causal SGA (Section \ref{sec:OW-Lasso}). Specifically, we treat the pre-specified subgroups as candidates for interactions with standard covariates in a logistic propensity score model and use LASSO to select important interactions. We then capitalize on the exact balance property of overlap weighting with a logistic regression to achieve covariate balance {\it both} overall and within subgroups, thus reducing bias and variance in causal SGA. Then, we show analytically that balancing covariates within a subgroup bounds the bias in estimating subgroup causal effects (Section \ref{sec:bias}). Finally, we device a new diagnostic graph, which we call the Connect-S plot, for visualizing the subgroup covariate balance (Section \ref{sec:connect4}). We conduct extensive simulation studies to compare the proposed method with several alternative methods (Section \ref{sec:sim}), and illustrate its application in a motivating example (Section \ref{sec:application}).

%It is worthwhile to point out the distinction between a recent stream of machine learning methods for HTE estimation and our method. Beside the aforementioned BART, tree methods such as the Causal Forest and Causal boosting, which are designed to identify the subpopulations with significant HTEs in a data-driven fashion. A main distinction between these methods  is the design- versus analysis-based approach to causal inference. The propensity score methods are design-based, which avoids modeling the outcome. In contrast, most of the above machine learning methods directly model the outcomes, bypassing the propensity scores and thus the balance issue. When the outcome model is correctly specified, such outcome-model-based method would produce valid and the most efficient estimates. However, the outcome model is almost always misspecified to some degree, threatening the validity of the causal conclusions.

%Under certain conditions, good balance is sufficient to eliminate bias in causal effect estimation regardless of whether the propensity score is correctly specified.  Thus, a number of approaches specify weights that do not necessarily involve a propensity score, but target balance of pre-specified covariates directly \citep{imai2014covariate,zubizarreta2015stable,lif18,zhao2019covariate}.

Our methodology is motivated from an observational comparative effectiveness study based on the Comparing Options for Management: Patient-centered Results for Uterine Fibroids (COMPARE-UF) registry. \citep{stewart2018comparing} Our goal  is to evaluate alternative management options for uterine fibroids for relief of symptoms and quality of life. Subgroup analysis was a primary aim to determine whether certain types of patient subgroups should receive myomectomy versus hysterectomy procedures. Investigators pre-specified 35 subgroups of interest based on categories of 16 variables including race, age, and baseline symptom severity. In addition, 20 covariates were considered as potential confounders, including certain demographics, disease history, quality of life and symptoms. The total sample size is 1430, with 567 patients in the myomectomy group and 863 patients in the hysterectomy group. There are in total 700 subgroup-confounder combinations, which pose great challenges to check and ensure balance for causal analyses.

\section{Estimands and estimation in causal subgroup analysis} \label{sec:estimand}
\subsection{Notation}
Consider a sample of $N$ individuals, where $N_1$ units belong to the treatment group, denoted by $Z = 1$, and $N_0$ to the control group, denoted by $Z=0$. We maintain the stable unit treatment value assumption (SUTVA), \citep{rubin1980randomization} which includes two sub-assumptions: there is (i) no different versions of the treatment (also known as consistency \cite{robins1986new}), and (ii) no interference between units. Under SUTVA, each unit $i$ has two potential outcomes $Y_i(1)$ and $Y_i(0)$ corresponding to the two possible treatment levels, of which only the one corresponding to the actual treatment assigned is observed, $Y_i = Z_i Y_i(1) + (1-Z_i)Y_i(0)$. We also observe a vector of $P$ pre-treatment covariates, $\bm{X}_i =(X_{i1}, ..., X_{iP})^T$.

We denote the subgroups of interest by indicator variables $\bm{S_i}=(S_{i1}, ..., S_{iR})^T$, where $S_{ir}=1$ if the $i^{th}$ unit is a member of the $r^{th} (r=1, 2, \dots,R)$ subgroup and 0 otherwise (e.g. Black race, male gender, and younger age).  Usually, $S_{ir} = f_{r}(\bm{X}_i)$ for some function  $f_{r}$ that defines categories based on $\bm{X}_i$.  The $R$ groups are not required to be mutually exclusive, and a unit $i$ can belong to multiple subgroups. In fact, we are particularly interested in one-at-a-time subgroup analysis where the groups compared are defined as $S_{ir} =0 $ and $S_{ir} =1$ for each $r$, while averaging over the levels of $\{S_{i1},..., S_{iR}\} \setminus \{S_{ir}\}$. Nonetheless, to simplify notation in Section \ref{swate}, we assume mutually exclusive subgroups so that $\sum_{r=1}^{R} S_{ir}=1$ hereafter.

The propensity score is $e(\bm{X}_i, \bm{S}_i) = \Pr(Z_i = 1| \bm{X}_i, \bm{S}_i)$.  When the components of $\bm{S}_i$ are functions of $\bm{X}_i$, the dependence of the propensity score on the subgroup indicators could be dropped.  However, the sub-grouping variables $\bm{S}_i$ may not all be a function of $\bm{X}_i$. Further, subgroups are most often defined based on physicians' and patients' prior knowledge with respect to which covariates are important for selecting treatment or with respect to the outcome.  For this reason the true propensity score may be subgroup-specific in that relationships between $\bm{X}_i$ and $Z_i$ depend on $\bm{S}_i$. For this reason, both the typical covariates $\bm{X}_i$ and the sub-grouping variables $\bm{S}_i$ are explicitly denoted. 

\subsection{The estimand: Subgroup weighted average treatment effect}
\label{swate}

Traditional causal inference methods focus on the average treatment effect (ATE), $\E_{f}[Y(1) - Y(0)]$, where the expectation is over the population with probability density $f(\bm x, \bm s)$ for the covariates and subgroups. Corresponding subgroup analysis would evaluate the subgroup average treatment effect (S-ATE), $\tau_r = \E_{f}[Y(1) - Y(0)|S_r = 1]$. Recently there has been increasing focus on weighted average treatment effects which represent average causal effects over a different, potentially more clinically relevant populations.\citep{crump2009dealing, lif18, tao2019doubly, zhao2019covariate, thomas2020using} We extend the weighted average treatment effect to the context of subgroup analysis.

Let $g(\bm x, \bm s)$ denote the covariate/subgroup density of the clinically relevant target population. The ratio $h(\bm x, \bm s)$=$g(\bm x, \bm s)/f(\bm x, \bm s)$ is called a \emph{tilting function}, \citep{li2019propensity} which re-weights the distribution of the observed sample to represent the target population. Denote the conditional expectation of the potential outcome in subgroup $r$ with treatment $z$ by $\mu_{rz}(\bm x)=\E_f\{Y(z)|\bm{X} = \bm x,S_{r} = 1 \}$ for $z=0,1$. Then, we can represent the subgroup weighted average treatment effect (S-WATE) over the target population by:
\begin{equation}
\tau_{r,h}=\E_g[Y(1)-Y(0)|S_{r} = 1]=\frac{\E\{h(\bm X, \bm S)(\mu_{r1}(\bm X)-\mu_{r0}(\bm X))|S_{r} = 1\}}{\E\{h(\bm X, \bm S)|S_{r} = 1\}}. \label{eq: swate}
\end{equation}

In practice, we specify the target population by pre-specifying the tilting function $h(\bm x, \bm s)$. Different choices of the function $h$ lead to different estimands of interest. For example, for $h(\bm x, \bm s) = 1$ the S-WATE collapses to the S-ATE: $\tau_{r,h} \equiv \tau_r$.  Another special case arises under homogeneity when $\mu_{r1}(\bm x)-\mu_{r0}(\bm x)$ is constant for all $\bm x$ and $\tau_{r,h} \equiv \tau_r$ for all $h$. Several common tilting functions will be discussed subsequently within the context of subgroup analysis.

To identify the S-WATE from observational data, we make two standard assumptions:\citep{rosenbaum1983central} (i) \textit{Unconfoundedness}: $Z \independent \{Y(1),Y(0)\}| \{\bm{X},\bm{S}\} $, which implies that the treatment assignment is randomized given the observed covariates, and (ii) \textit{Overlap (or positivity)}:  $0<e(\bm{X}_i, \bm{S}_i)<1$, which requires that each unit has a non-zero probability of being assigned to either treatment condition. Then, we can estimate the S-WATE in subgroup $r$, $\tau_{r,h}$, using the H\'ajek estimator 
\begin{equation}
\widehat \tau_{r,h} = \frac{\sum_{i = 1}^N Z_i S_{ir} w_{i1} Y_i}{\sum_{i = 1}^N Z_i S_{ir} w_{i1}} 
-  \frac{\sum_{i = 1}^N(1 - Z_i) S_{ir} w_{i0} Y_i}{\sum_{i = 1}^N(1 - Z_i) S_{ir} w_{i0}},
\label{trh}
\end{equation}
where the weights $w$ are the balancing weights corresponding to the specific tilting function $h(\bm x, \bm s)$ (equivalently the target population $g(\bm x, \bm s)$):\cite{lif18} 
\begin{equation}
\begin{cases}
w_{i1} = \dfrac{h(\bm{X}_i, \bm{S}_i)}{e(\bm{X}_i, \bm{S}_i)} & \text{for}~Z_i=1,\\
w_{i0} = \dfrac{h(\bm{X}_i, \bm{S}_i)}{1-e(\bm{X}_i, \bm{S}_i)} & \text{for}~Z_i=0.
\end{cases}\label{weight}
\end{equation}
The most widely used balancing weights are the inverse probability weights (IPW),\cite{Robins2000} $(w_1 = 1/e(\bm x, \bm s), w_0=1/(1-e(\bm x, \bm s))$, corresponding to $h(\bm x, \bm s)=1$. The target population of IPW is the combination of treated and control patients that are represented by the study sample, and the subgroup-specific estimand is the subgroup average treatment effect (S-ATE). The balancing weights which will play a key role in this paper (Sections \ref{sec:exactbalance} and \ref{sec:OW-Lasso}) are the overlap weights (OW), ($w_1 = 1-e(\bm x, \bm s), w_0 = e(\bm x, \bm s)$), corresponding to $h(\bm x, \bm s)= e(\bm x, \bm s)(1-e(\bm x, \bm s))$.\cite{lif18} Balancing weights are defined on the entire sample and are applicable to subgroups where the value of $\bm{S}_i$ is fixed and defines the subgroup of interest. We show in the Web Appendix 1.1 that $\widehat \tau_{r,h}$ is consistent for $\tau_{r,h}$.

In practice, the propensity score, $e(\bm{X}_i, \bm{S}_i)$, is usually not known and is estimated from the data. Then, the weights $w_{i}$ in \cref{trh} are replaced with $\widehat{w}_{i}$ based on the estimated propensity score $\widehat e(\bm{X}_i, \bm{S}_i)$.  While balancing the true propensity score would balance the covariates in all covariate-defined subgroups in expectation, the estimated weights $\widehat{w}_{i}$ based on an estimated propensity score often fail to achieve covariate balance, particularly within subgroups.\citep{dong2019subgroup} As we show in Section \ref{sec:bias}, covariate balance in the subgroups is crucial for unbiased estimation of the S-WATE. Therefore, it is beneficial to choose weights that guarantee balance. 

%In Section \ref{sec:OW-Lasso} we adapt the overlap weights for this purpose.  Before we dive into the question of \emph{how to balance}, we first need address \emph{what to balance}. Specifically, it is necessary to first consider what functions of covariates (e.g. moments) should be balanced in estimating subgroup ATEs and how this differs from estimation of the overall ATE. We address this question in the next Section.  

%\section{Estimation of subgroup causal effects} \label{sec:estimation}
\subsection{Exact subgroup balance via overlap weights} \label{sec:exactbalance}
%Overlap weights are where $h(\bm x, \bm s) = e(\bm x, \bm s)(1-e(\bm x, \bm s))$ in equation \eqref{weight} and $\tau_{r,h}$ is the subgroup average treatment effect in the overlap population (S-ATO).\citep{lif18} 

We propose to use overlap weighting to achieve exact balance on the subgroup-specific covariate means. As noted above, the overlap weight of each unit is the probability of being assigned to the opposite group: $w_1 = 1-e(\bm x, \bm s)$ and $w_0 = e(\bm x, \bm s)$, arising from tilting function $h(\bm x, \bm s)= e(\bm x, \bm s)(1-e(\bm x, \bm s))$. This tilting function is maximized for individuals with propensity scores close to 0.5, i.e. those who are equally likely to be treated or not, and minimized for individuals with propensity scores close to 0 or 1, i.e. those who are nearly always treated or never treated. Consequently, the target population of OW emphasizes covariate profiles with the most overlap between treatment groups and the subgroup-specific estimand is the subgroup average treatment effect of the overlap population (S-ATO). Though statistically defined, this represents a target population of intrinsic substantive interest. \citep{lif18,li2019addressing, thomas2020overlap} Specifically, the overlap population mimics the characteristics of a pragmatic randomized trial that is highly inclusive, excluding no study participants from the available sample, but emphasizing the comparison of patients at clinical equipoise. The resulting target population can be empirically described through a weighted baseline characteristics table. When the S-ATO is clinically relevant, its corresponding weighting estimator has attractive properties regarding balance and variance.

First, OWs have a unique finite-sample property of exact balance. Specifically, outside the context of SGA, Li et al.\cite{lif18} show that when the propensity score is estimated by a logistic regression, overlap weighting leads to exact balance on the weighted covariate means. We extend this property to subgroups as follows.

%%%%%%%%%%%%%%%%%%%%%%   prop 3   %%%%%%%%%%%%%%%%%%%%%%%%%%%%%%%%%%

\begin{prop}
If the postulated propensity score model is logistic regression with subgroup-covariate interactions, i.e. %$\hat{e}(\bm{X}_i, \bm{S}_i)= \text{logit}^{-1} (\hat{\beta}_0  + \bm{X}_i^T \bm{\hat{\beta}_x}  + \bm{S}_i^T \bm{\hat{\beta}_s}   + (\bm{X}_i \cdot \bm{S}_i)^T \bm{\hat{\beta}_{xs}})$, where $\bm{\hat{\beta}} = (\hat{\beta}_0, \bm{\hat{\beta}_x}^T,  \bm{\hat{\beta}_s}^T, \bm{\hat{\beta}_{xs}}^T)^T$ 
$\hat{e}(\bm{X}_i, \bm{S}_i)= \text{logit}^{-1} (\hat{\alpha}_0  + \bm{X}_i^T \bm{\hat{\alpha}_x}  + \bm{S}_i^T \bm{\hat{\alpha}_s}   + (\bm{X}_i \cdot \bm{S}_i)^T \bm{\hat{\alpha}_{xs}})$, where $\bm{\hat{\alpha}} = (\hat{\alpha}_0, \bm{\hat{\alpha}_x}^T,  \bm{\hat{\alpha}_s}^T, \bm{\hat{\alpha}_{xs}}^T)^T$
is the maximum likelihood (ML) estimator and $(\bm{X}_i \cdot \bm{S}_i)$ denotes all pairwise interactions between $\bm{X}_i$ and $\bm{S}_i$, then the overlap weights lead to exact mean balance in the subgroups and overall:
\[\sum_{i=1}^N  Z_i S_{ir} X_{ip} \hat{w}_{i1} -
\sum_{i=1}^N  (1-Z_i)S_{ir} X_{ip}\hat{w}_{i0} = 0, \quad
\text{for all } r = 1, 2, \dots, R, \text{ and } p = 1, 2, \dots, P. \]
Again the weights need to be normalized such that $\sum_i^N  Z_i S_{ir} \hat{w}_{i1} =
\sum_i^N  (1-Z_i) S_{ir} \hat{w}_{i0} = 1$ (Web Appendix 1.5).
\label{prop:ps_balance}
\end{prop}
 
Proposition \ref{prop:ps_balance} implies that when a logistic model for propensity scores is augmented to include $(\bm{X}_i \cdot \bm{S}_i)$ and paired with OW, exact balance is achieved \emph{both overall and within subgroups}.  
Additionally, the approach can be motivated by focusing on correct specification of the propensity score model in the scientific context.  When subgroups are defined \emph{a priori} it is usually based on clinical knowledge of which patient characteristics are most likely to alter the treatment effect.  Thus treatment decisions in the observational data may already be different in these subgroups, corresponding to covariate-subgroup interactions in the true propensity score model. This motivates the inclusion of pre-specified subgroups as candidates for interactions with standard covariates in the propensity score model. However, as the propensity score model approaches saturation, the estimated propensity scores will converge to 0 and 1, thus causing variance inflation in the treatment effect estimates. 

Variance inflation with increasing PS model complexity is partially mitigated by OW.  OWs are naturally bounded between 0 and 1, thus can avoid the issues of extreme weights and large variability that can occur when $h(\bm x, \bm s) = 1$.\citep{lif18}  In fact, the overlap tilting function $h(\bm x, \bm s) = e(\bm x, \bm s)(1-e(\bm x, \bm s))$ gives the smallest large-sample variance of the weighted estimator $\widehat \tau_{r,h}$ over all possible $h$ under homoscedasticity (Web Appendix 1.1). For subgroup analysis, the optimal efficiency helps to mitigate the potential variance inflation arising from a more complex propensity score model. Nonetheless, when the number of covariates and/or subgroups is large, variable selection in the propensity score model is necessary. Therefore, we propose a new method for causal SGA to accommodate considerations on both variable selection and covariate balance.   

\subsection{Combining overlap weights with post-LASSO for causal SGA: the OW-pLASSO algorithm} \label{sec:OW-Lasso}
We propose the \emph{OW-pLASSO} algorithm for causal SGA, which combines two main components. The first component uses the post-LASSO approach \citep{belloni13, james2013introduction} to select covariate-subgroup interactions and estimate the propensity scores. In causal settings regularization inadvertently biases treatment effect estimates by over-shrinking regression coefficients. \citep{hahn2018regularization} Hence, we adopt the post-LASSO approach instead of the original LASSO \citep{tibshirani1996regression}. The second component uses OW to achieve covariate balance in the subgroups.  

The \emph{OW-pLASSO} algorithm consists of the following steps:
\emph{
\begin{enumerate}
\item[S1.] Fit a logistic propensity score model with all pre-specified covariates and subgroup variables along with pairwise covariate-subgroup interactions, i.e design matrix $(\bm{X}_i, \bm{S}_i, \bm{X}_i \cdot \bm{S}_i)$, and perform LASSO to select covariate-subgroup interactions (without penalizing the main effects in the model).
\item[S2.] Estimate the propensity scores by refitting the logistic regression with all main effects and selected covariate-subgroup interactions from S1.
\item[S3.] Calculate the overlap weights based on the propensity scores estimated from S2, and check subgroup balance using the Connect-S plot (Section \ref{sec:connect4}) before and after weighting.
\item[S4.] Estimate the causal effects for all prespecified subgroups using Estimator \eqref{trh} with the overlap weights from S3.
\end{enumerate}
}
From extensive simulation studies (Section \ref{sec:sim}), we find the the \emph{OW-pLASSO} algorithm outperforms combinations of IPW and other popular machine learning models for propensity scores in estimating the S-WATE estimands. One of the key reasons of OW-pLASSO's advantage is that it achieves within-subgroup exact mean balance, which is crucial for bias reduction, as we show analytically in the next Section. 

To estimate the variance of the (overall and subgroup) treatment effects, we suggest two methods: (i) the robust sandwich estimator, as recently described for IPW; \citep{austin2017performance} this approach is known to be slightly conservative as it does not take into account the uncertainty in estimating the propensity scores, but has been shown to work well in practice. (ii) Bootstrapping: estimate propensity scores in the original sample using post-LASSO and treat the estimated propensity scores as fixed when estimating the causal effects in each bootstrap sample. Note that in the bootstrap method, we caution against the practice of re-fitting the propensity score using post-LASSO in each bootstrap sample, since the bootstrap is {\it in}consistent for LASSO estimators. \cite{chatterjee2010asymptotic}
Since uncertainty for LASSO estimators is hard to quantify, none of the variance estimation approaches we consider aims to incorporate the variability of propensity score estimates. However, ignoring the uncertainty of the propensity score is justifiable in causal inference studies\cite{Hirano2003} as the propensity score is often viewed as part of the ``design'' phase of a study.\cite{ho2007matching}
Our simulation studies in Web Appendix 2.3 validate these variance estimation methods coupled with the proposed OW-pLASSO algorithm. 

\section{Bounding bias for subgroup causal effects} 
\label{sec:bias}
When focusing on additive models, Zubizarreta \cite{zubizarreta2015stable} showed that the weighting estimator for the population mean is unbiased when the covariate means are balanced. We extend this work to subgroup analysis by showing that balance of covariates within a subgroup leads to minimal bias of the estimator $\widehat \tau_{r, h}$. In \cref{prop:balance_bias}, we show this result when the treatment effect is homogeneous within a subgroup ($\tau_{r,h} = \tau_r$), and in \cref{prop:balance_hte} we extend it to allow for within-subgroup effect heterogeneity. In both cases, treatment effects are allowed to vary between subgroup levels. 

\begin{prop}  Suppose that the outcome surface satisfies an additive model, e.g.
$Y_i(z) = \sum_{r = 1}^R \beta_r S_{ir} + $ $\sum_{r = 1}^R \sum_{p = 1}^P \beta_{rp} S_{ir}X_{ip} +\sum_{r = 1}^R \tau_r S_{ir} z +$ $\epsilon_i(z),$ with $\E[\epsilon_i(z) | \bm{X}_i, \bm{S}_i] = 0$.  For any weight $w_{i}$ that is normalized within subgroups (i.e. $\sum_{i = 1}^N Z_i S_{ir} w_{i1} =\sum_{i = 1}^N (1 - Z_i) S_{ir} w_{i0}=1$), if mean balance holds in the $r^{th}$ subgroup, expressed as
\begin{equation}
\left| \sum_{i = 1}^N Z_i S_{ir} w_{i1} X_{ip} - \sum_{i = 1}^N (1 - Z_i) S_{ir} w_{i0}  X_{ip} \right| < \delta, \ \ \text{for all } p = 1, 2, \dots, P,
\label{mean_condition}
\end{equation}
then the bias for the $r^{th}$ subgroup is bounded,  $\left| E \left[\widehat \tau_{r,h} - \tau_{r} \right] \right|  < \delta \sum_{p = 1}^P |\beta_{rp}|$ (Web Appendix 1.2).
\label{prop:balance_bias}
\end{prop}
Therefore, any weighting scheme for which $\delta \approx 0$ will eliminate bias for SGA when the outcome satisfies an additive model.  \cref{prop:balance_bias} illustrates that mean balance in the overall sample, $\left| \sum_{i = 1}^N Z_i w_{i1} X_{ip} - \sum_{i = 1}^N (1 - Z_i)  w_{i0}  X_{ip} \right| < \delta$, is {\it not} sufficient, and balance is required {\it within the subgroup}.  Even in the special case where the true response surface is additive in the covariates and the treatment effect is constant ($\beta_{rp} = \beta_p$, and $\tau_r = \tau)$, the subgroup-specific Condition (\ref{mean_condition}) is still necessary to ensure minimal bias of $\widehat \tau_{r,h}$.   
%%%%%%%%%%%%%%%%%%%%%%%%%%%%%%%%%%%%%%%%%%%%%%%%%%%%%%%%%%%%%%%%%%%%

%%%%%%%%%%%%%%%%%%%%%%   prop 2   %%%%%%%%%%%%%%%%%%%%%%%%%%%%%%%%%%

\begin{prop}  Suppose the additive model is relaxed to allow treatment effect heterogeneity by covariates $\bm{X}_i$ within subgroups: $Y_i(z) = \sum_{r = 1}^R \beta_r S_{ir} + $ $\sum_{r = 1}^R \sum_{p = 1}^P \beta_{rp} S_{ir}X_{ip} +$
$\sum_{r = 1}^R \tau_r S_{ir} z +$ $ \sum_{p = 1}^P \gamma_{rp} S_{ir} X_{ip} z +$ $\epsilon_i(z),$ with $E[\epsilon_i(z) | \bm{X}_i, \bm{S}_i] = 0$.  If Condition (\ref{mean_condition}) holds and additionally,
\begin{equation}
\left| \sum_{i = 1}^N Z_i S_{ir} w_i X_{ip} - \dfrac{\sum_{i = 1}^N h(\bm{X}_i, \bm{S}_i) S_{ir} X_{ip}}{\sum_{i = 1}^N h(\bm{X}_i, \bm{S}_i) S_{ir}}  \right| < \delta_{2}, \ \ \text{for all } p = 1, 2, \dots, P,
\label{dist_condition}
\end{equation}
then the bias for the $r^{th}$ subgroup is bounded, $ \left| E \left[\widehat \tau_{r,h} - \tau_{r,h} \right] \right|  < \delta \sum_{p = 1}^P |\beta_{rp}| + \delta_{2} \sum_{p = 1}^P |\gamma_{rp}|$ (Web Appendix 1.3).
\label{prop:balance_hte}
\end{prop}
Condition \eqref{dist_condition} requires the weighted sample covariate mean of treated patients within the subgroup to be close to the subgroup target population covariate mean.  This condition can be verified when $h$ is a pre-defined function, but not when $h(\bm{X}_i, \bm{S}_i)$ depends on an unknown propensity score $e(\bm{X}_i, \bm{S}_i)$.  However this term is expected to be small unless the model for the propensity score is severely mis-specified.  In the Web Appendix 1.4, we show that an alternative, verifiable condition: $\left| \sum_{i = 1}^N Z_i S_{ir} w_i X_{ip} - \dfrac{\sum_{i = 1}^N \widehat h(\bm{X}_i, \bm{S}_i) S_{ir} X_{ip}}{\sum_{i = 1}^N \widehat h(\bm{X}_i, \bm{S}_i) S_{ir}}  \right| < \delta_{2}$, is sufficient if we are willing to estimate a slightly different estimand, namely, the subgroup-sample weighted average treatment effect (S-SWATE), $\tau_{r,\widehat h} =\dfrac{\sum_i  \widehat h(\bm{X}_i, \bm{S}_i)[\mu_{r1}(\bm X_i, \bm S_i)-\mu_{r0}(\bm X_i, \bm S_i)]S_{ir}}{\sum_i  \widehat h(\bm{X}_i, \bm{S}_i)S_{ir} }$.  Therefore, verifiable mean balance conditions are sufficient for $ \widehat \tau_{r, h}$ to have a causal interpretation, but the propensity score model must be approximately correct in order for the weighted population to correspond to the target population and estimate $\tau_{r,h}$. Similarly to Condition \eqref{mean_condition}, Condition \eqref{dist_condition} can be checked by the Connect-S plot (Section \ref{sec:connect4}).

It is instructive to consider the special case were $h(\bm{X}_i, \bm{S}_i)=1$ and the target population is the sampled population.  In this case, $h$ is known and Condition \eqref{dist_condition} can be empirically verified.  However, it will not necessarily be satisfied for weights based on an estimated propensity score. To the best of our knowledge, Condition \eqref{mean_condition} is typically checked but Condition \eqref{dist_condition} is not. Under heterogeneous treatment effects this second condition is needed. This reveals a potential risk of using weights that balance covariates without defining a tilting function and target estimand (S-WATE). \citep{imai2014covariate,zubizarreta2015stable,lif18,zhao2019covariate} The implicit estimand is the S-ATE with $h(\bm{X}_i, \bm{S}_i)=1$.  While these methods are designed to satisfy Condition \eqref{mean_condition}, Condition \eqref{dist_condition} does not play a role in the construction of the weights and may be violated. 

The assumption of linearity in the covariates can be relaxed and the non-linear case is addressed in Web Appendix 1.6 (Proposition 4). We find that mean balance remains an important condition for unbiasedness, but various higher order moments are potentially important, depending on the true model. Whether it would be practically feasible to pre-specify and interpret the corresponding, higher order balance checks, particularly in finite samples, requires future investigation. We do not undertake that here, but instead focus on correct estimation of the propensity score model, coupled with mean balance which is sufficient in linear models (above) and necessary in non-linear models.

\section{Visualizing subgroup balance: The Connect-S plot} 
\label{sec:connect4}
In practice, it is often difficult to assess whether existing propensity score methods achieve the balance conditions defined in Section \eqref{sec:bias}. For example, in the motivating application of COMPARE-UF, there are 700 combinations of subgroups and covariates for which to check Condition \eqref{mean_condition}.  In this Section we introduce a new graph for visualizing subgroup balance -- the Connect-S plot. We first introduce two important metrics that will be presented in the plot.   

\begin{figure}[ht]

\centering\includegraphics[width=14cm]{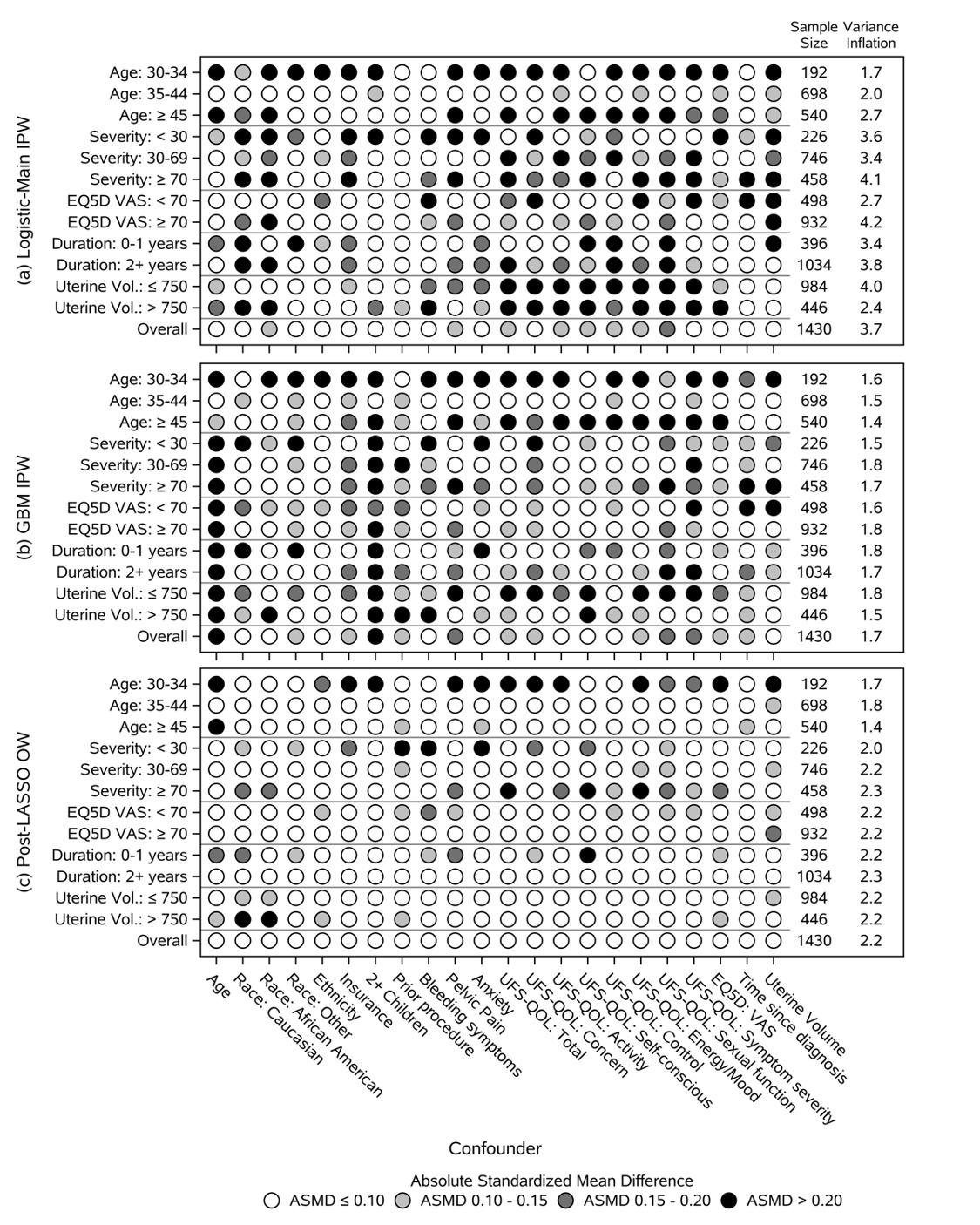} 
\caption[]{The Connect-S plot of the subgroup ASMD and approximate variance inflation in COMPARE-UF after applying balancing weights for adjustment by (a) Logistic-Main IPW, propensity score estimated by main effects logistic regression with IPW; (b) GBM IPW, propensity score estimated by generalized boosted models with IPW; (c) OW-pLASSO, propensity score estimated by post-LASSO with OW. Select subgroups are displayed in rows and all confounders are displayed in columns. }
\label{fig:ConnectS}
\end{figure}

The first statistic is the \emph{absolute standardized mean difference} (ASMD), \cite{austin2015moving} which is widely used for measuring covariate balance. The ASMD is the difference in weighted means, defined in Condition \eqref{mean_condition}, further scaled by the pooled, unweighted standard deviation.  That is
\begin{equation}
\text{ASMD}_{r,p} = \frac{ \sum_{i = 1}^N Z_i S_{ir} w_{i1} X_{ip} - \sum_{i = 1}^N (1 - Z_i) S_{ir} w_{i0}  X_{ip} }{s_{r,p}}
 \label{asmd}
 \end{equation}
where $s_{r,p}$ is the unweighted, pooled standard deviation for the  $r^{th}$ subgroup and the $p^{th}$ covariate (See Web Appendix 1.5 for details).  Scaling by $s_{r,p}$ facilitates a practical interpretation of the weighted mean difference, relative to the standard deviation of the variable $X_p$. Various rules of thumb suggest that the $\text{ASMD}_{r,p}$ should be less than 0.10 or 0.20 (i.e. an acceptable $\delta$ is <0.10 to 0.20). \cite{austin2015moving}

% Move to section in the Appendix
% I think for completeness we should have the definition in our paper. 
% Rather than people ahve to go to another reference. But it is ugly. 
% Consider appendix. 
%Define: s_{r,p}=\sqrt{{s_{r,p,0}^2}/{2} + {s_{r,p,1}^2}/{2}}
%Following the literature, \cite{austin2015moving} we define the subgroup weighted mean within arm %$z= 0, 1$ as $\bar{X}^w_{p,1}={\sum_i  Z_i w_i S_{ir}X_{ip}}/{\sum_i Z_i w_i S_{ir}}$ and %$\bar{X}^w_{p,0}={\sum_i  (1-Z_i) w_i S_{ir}X_{ip}}/{\sum_i (1-Z_i) w_i S_{ir}},$ respectively,
%and the subgroup weighted variance of $X_p$ within arm $z= 0, 1$ as 
%\begin{eqnarray*}
%s_{r,p,1}^2&=&\frac{\sum_i Z_i w_i S_{ir}}{(\sum_i Z_i w_i S_{ir})^2-\sum_i (Z_i w_i %S_{ir})^2}\sum_iZ_i w_i S_{ir}(X_{ip}-\bar{X}^w_{p,1})^2,
%\end{eqnarray*}
%and
%\begin{eqnarray*}
%s_{r,p,0}^2&=&\frac{\sum_i (1-Z_i) w_i S_{ir}}{(\sum_i (1-Z_i) w_i S_{ir})^2-\sum_i ((1-Z_i) w_i %S_{ir})^2}\sum_i(1-Z_i) w_i S_{ir}(X_{ip}-\bar{X}^w_{p,0})^2.
%\end{eqnarray*}

The second metric concerns variance. In the context of SGA, the propensity score model is typically complex, including many interaction terms. Therefore, a particularly important consideration in propensity score weighting is the variance inflation due to model complexity.  Li et al. \cite{lif18} suggested to use the following statistic akin to the ``design effect" approximation of Kish \cite{kish1965survey} in survey literature to approximate the \emph{variance inflation} (VI): 
\begin{equation}
\mbox{VI}=(1/N_1+1/N_0)^{-1}\sum_{z=0,1}\frac{\sum_{i=1}^{N_z}w_{iz}^2}{\left(\sum_{i=1}^{N_z}w_{iz}\right)^2}, \label{eq:varianceinflation}
\end{equation}
where $N_z$ is the sample size of treatment group $z$. For the unadjusted estimator, $w_{iz}=1$ for all units and VI=1. Increasing values of VI imply increasingly worse efficiency for alternative weighting algorithms.  It is straightforward to define the subgroup-specific version of the variance inflation statistic.  

The Connect-S plot for $S$ subgroups resembles the rectangular grid of a Connect4 game: each row represents a subgroup variable, (e.g. a race group), and the name and subgroup sample size is displayed at the beginning and the end of each row, respectively; each column represents a confounder that we want to balance (e.g. age). Therefore, each dot corresponds to a specific subgroup $S$ and confounder $X$, and the shade of the dot is coded based on the ASMD of confounder $X$ in subgroup $S$, with darker color meaning more severe imbalance. The end of each row also presents subgroup-specific approximate variance inflation.

Panel (a) of Figure \ref{fig:ConnectS} presents the Connect-S plot for COMPARE-UF after adjustment by IPW where the propensity score for myomectomy versus hysterectomy is estimated by main effects logistic regression. The bottom row of this panel shows that this method does a good job of balancing the confounders, overall. However, it does a poor job of achieving balance within subgroups. For example, subgroups based on age, symptom severity, EQ5D quality of life score, and uterine volume have many ASMDs greater than 0.10 and often greater than 0.20. These are not generally acceptable and motivate alternative methodology. A potential solution would be to use a more flexible model for the propensity score that does not assume main effects. Panel (b) of Figure \ref{fig:ConnectS} shows that balance in COMPARE-UF is not improved by estimating the propensity score with generalized boosted models and results were similar for random forest and BART methods (Web Appendix 2.4).

\section{Simulations} \label{sec:sim}
We compare the proposed OW-pLASSO method with a number of popular machine learning propensity score methods via simulations under different levels of confounding, sparsity and heterogeneity in causal SGA. 
\subsection{Simulation Design}
\noindent \emph{Data Generating Process.} In alignment with the COMPARE-UF study we generate $N=3000$ patients, with $P \in \{18, 48\}$ independent covariates $\bm{X}_i$, half of which drawn from a standard normal distribution $N(0,1)$, and the other half from Bernoulli(0.3). Two subgroup variables $ \bm{S}_i= (S_{i1}, S_{i2})$ are independently drawn from Bernoulli(0.25). The treatment indicator $Z_i$ is generated from Bernoulli($e(\bm{X}_i, \bm{S_i})$), with the \emph{true propensity score model}:
\begin{equation}
\text{logit}(e(\bm{X}_i,\bm{S}_i))= \alpha_r +\bm{S}_i^T \bm{\alpha}_s  +\bm{X}_i^T \bm{\alpha}_x   + (\bm{X}_i \cdot \bm{S}_i)^T \bm{\alpha_{xs}},
\label{inter_ps}
\end{equation}
with coefficients $\bm{\alpha} = (\alpha_r,\bm{\alpha_s}^T, \bm{\alpha_x}^T,  \bm{\alpha_{xs}}^T)^T$.

We set the coefficients in model (\ref{inter_ps}) as follows: $\alpha_r = -2$, $\bm{\alpha}_{s}^T = (1, 1)$. Out of the $P$ coefficients in $\bm{\alpha}_x$, $\psi$ portion of them have nonzero coefficients (i.e. true confounders in our simulation). The coefficients for the continuous and binary confounders take equally distanced values between $(0.25\gamma, 0.5\gamma)$, separately, and the rest are zeros. Last, we set $\bm{\alpha}_{xs} = -\bm{\alpha}_x \kappa$. To create a range of realistic scenarios in SGA we vary the three hyperparameters $(\psi, \gamma,\kappa)$ in the true propensity score model: 1) $\psi \in \{0.25, 0.75\}$ controls the proportion of covariates $\bm{X}_i$ that are true confounders; 2)  $\gamma \in  \{1, 1.25, 1.5\}$ controls the scale of the regression coefficients for $\bm{X}_i$, and 3) $\kappa\in \{0.25, 0.5, 0.75\}$ scales the regression coefficients for $(\bm{X}_i \cdot \bm{S}_i)$. For example, for $P=18, \gamma=1, \psi=0.25$, and $ \kappa=0.5$, the above setting specifies $\bm\alpha_x^T=(0.25,0.5,\bm 0_7, 0.25,0.5,\bm 0_7), \bm\alpha_{xs}^T=(-0.125,-0.25,\bm 0_7, -0.125,-0.25,\bm 0_7)$, where  $\bm 0_k$  is a k-vector of zeros. The above simulation settings mimic a common SGA situation in clinical studies. Specifically, when $S_1=1, S_2=1$, the two subgroup variables represent high risk conditions associated with the outcome  (e.g. risk score) and increase the likelihood of being treated. In the presence of these high risk conditions, other patient characteristics $\bm{X}_i$ play a lesser role in driving treatment decisions; this is reflected by the fact that magnitude of $\bm{\alpha}_x$ in the propensity model is smaller than $\bm{\alpha}_s$.  In the Web Appendix 2.1 we show that these specifications lead to treated and control units with various amounts of overlap for the true propensity score distributions. 

Next, a continuous outcome $Y_i$ (e.g. risk score) is generated from a linear regression model:
\begin{equation}
Y_i  =\beta_0 + \bm{X}_i^T \bm{\beta}_x     + \bm{S}_i^T \bm{\beta}_{s}   + \beta_{z}  Z_i  + (\bm{S}_i \cdot Z_i)^T \bm{\beta}_{sz}    + \epsilon_i,
\end{equation}
where $(\bm{S}_i \cdot Z_i)$ is a vector of all possible interactions between subgroup variables and treatment assignment, and $\epsilon_i$ is independently sampled from $N(0,1)$. We fix the model parameter $\beta_0 = 0$, $\bm{\beta}_x = \bm{\alpha}_x$, $\bm{\beta}_{s}^T = (0.8, 0.8)$, $\beta_z = -1$, and vary $\bm{\beta}_{sz}^T=(\beta_{1z},\beta_{2z} )^T \in \{(0,0), (0.5, 0.5
)\}$. When $\bm{\beta}_{sz}^T = (0,0)$, the treatment effect is homogeneous, and $\tau_r=\beta_z=-1$ for all subgroups. When $\bm{\beta}_{sz}^T = (0.5,0.5)$, the underlying treatment effect is heterogeneous within subgroups and between different subgroup levels. For example, when $P=18, \psi=0.25,\gamma=1, \kappa=0.75$, the true causal effect $\tau_h=-0.67$ for ATO, and $-0.75$ for ATE; $\tau_{\{S_{1}=0,h\}} = \tau_{\{S_{2}=0,h\}} = -0.83$ for S-ATO, and $-0.87$ for S-ATE; $\tau_{\{S_{1}=1,h\}} = \tau_{\{S_{2}=1,h\}} = -0.35$ for S-ATO, and $-0.37$ for S-ATE.

%$\tau_{r,h}=\tau_r+\sum_{p = 1}^P \gamma_{rp}\frac{\E[h(\bm{X},\bm{S})X_p|S_{r}=1]}{\E[h(\bm{X},\bm{S})|S_{r}=1]}$. ). Similar to the PS model, patients' other characteristics $\bm{X}_i$ are less prognostic of the outcome in the presence of high risk factors  $\bm{S}_i$ (strength of $\bm{\beta}_x$ in the outcome model is less than $\bm{\beta}_s$).

%\subsection{Alternative Methods}
%\label{s:psmodel}

\emph{Postulated propensity score models.} To estimate the propensity scores, we compare Post-LASSO with several popular alternatives in the literature: %\citep{breiman2001random,buhlmann2003boosting,mccaffrey04, chipman2010bart, hill2011challenges, wager2018estimation} : 
(1) True model: Logistic regression fitted via maximum likelihood (ML) with the correctly specified propensity score \eqref{inter_ps}, representing the oracle reference; (2) Logistic-Main: logistic regression with only main effects of the predictors  $(\bm{X}_i, \bm{S}_i)$ fitted via ML, representing the standard practice; (3) LASSO: LASSO\cite{tibshirani1996regression}  with the design matrix $(\bm{X}_i, \bm{S}_i, \bm{X}_i \cdot \bm{S}_i)$, implemented by the R package \textit{glmnet} without penalizing the main effects, and ten-fold cross validation for hyperparameter tuning;\citep{glmnet} (4) Post-LASSO: Logistic  regression  model fitted via ML  with  the covariate-subgroup interactions  selected  from  the  preceding LASSO;\cite{belloni2013least} (5) RF-Main: Random Forest (RF) \citep{breiman2001random, wager2018estimation} with the design matrix $(\bm{X}_i, \bm{S}_i)$, implemented by R package \textit{ranger} with default hyperparameters and 1000 trees;\citep{wright15} (6)  RF-All: RF with the augmented design matrix $(\bm{X}_i, \bm{S}_i, \bm{X}_i \cdot \bm{S}_i)$;  Among the examined scenarios, we observe no difference between the RF-All and RF-Main PS model, suggesting that RFs performance depends little on the provided design matrix.  For simplicity, we omit results on RF-All; (7) GBM: Generalized boosted model (GBM) \citep{buhlmann2003boosting,mccaffrey04} with the design matrix $(\bm{X}_i, \bm{S}_i)$, implemented by R package \textit{twang} with 5000 trees, interaction depth equals to 2, and other default hyperparameters;\citep{ridgeway2008twang} (8) BART: Bayesian additive regression  trees \citep{chipman2010bart} with the design matrix $(\bm{X}_i, \bm{S}_i)$, using the R function \textit{pbart} in package \textit{BART} with default hyperparameters.

Each of the preceding propensity score models is paired with (a) inverse probability of treatment weighting (IPW) and (b) overlap weighting (OW).  All the simulation analyses are conducted under R version 3.4.4. In total, there are 72 scenarios examined by the factorial design, with 100 replicate data sets generated per scenario. 

%\subsection{Performance Assessment}

\emph{Performance metrics.} The performance of different approaches is compared overall (averaged over subgroups) and within four subgroups defined by $S_{i1}=0$, $S_{i1}=1$, $S_{i2}=0$, $S_{i2}=1$. First, we check balance of covariates by the ASMD of each covariate, averaged across the 100 simulated data sets, and calculate the maximum ASMD value across all covariates. Second, we consider the relative bias and root mean squared error (RMSE) to study the precision and stability of various estimators.

\subsection{Simulation Results}
\begin{figure}[!b]
\centering
    \includegraphics[width=6in]{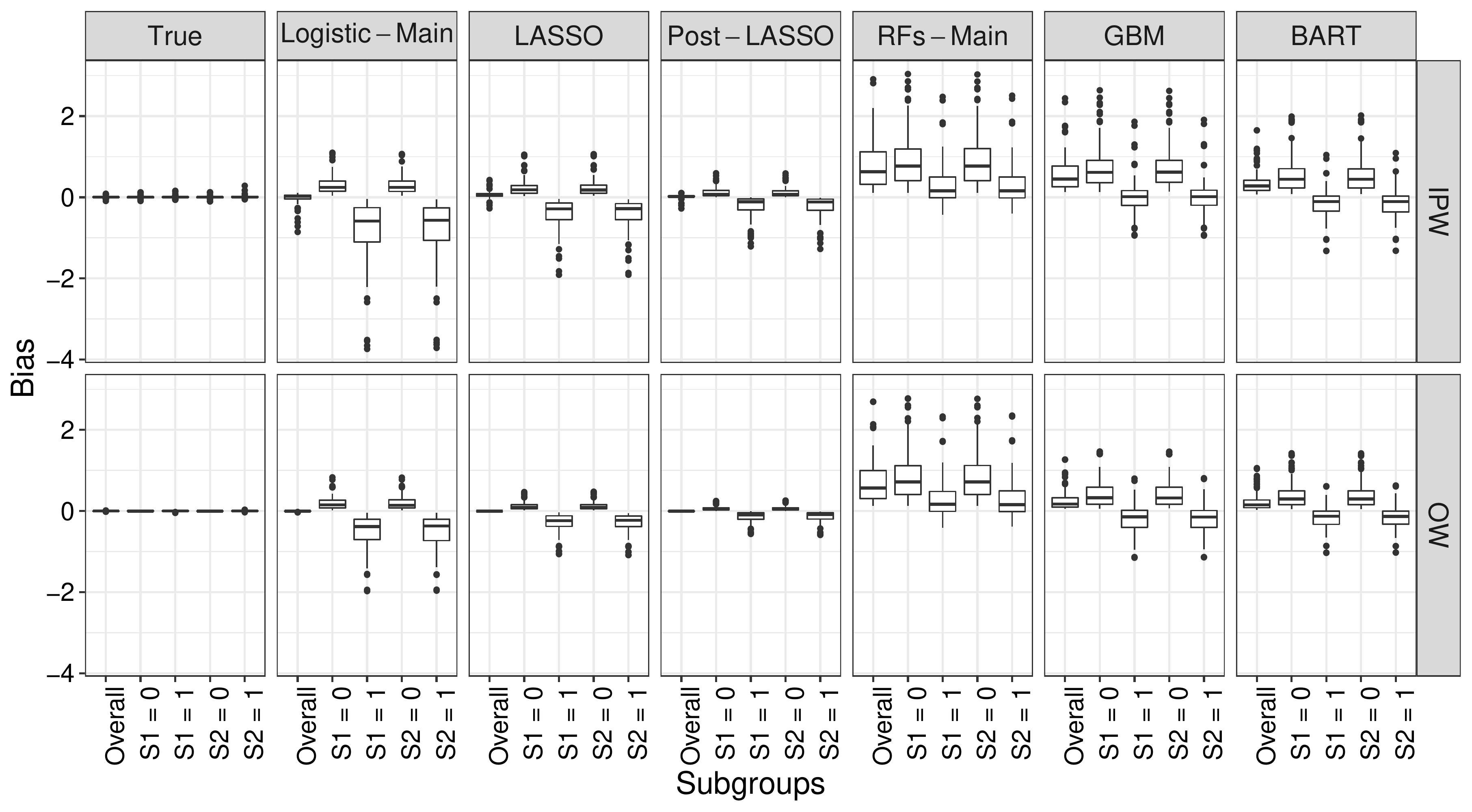}
    \caption{Bias in estimating the overall WATE and the four subgroup S-WATE across different postulated propensity models and weighting schemes. Each dot represents one of the 72 simulation scenarios.}
    \label{bias}
\end{figure}

\begin{figure}[ht]
    \centering
    \includegraphics[width=6in]{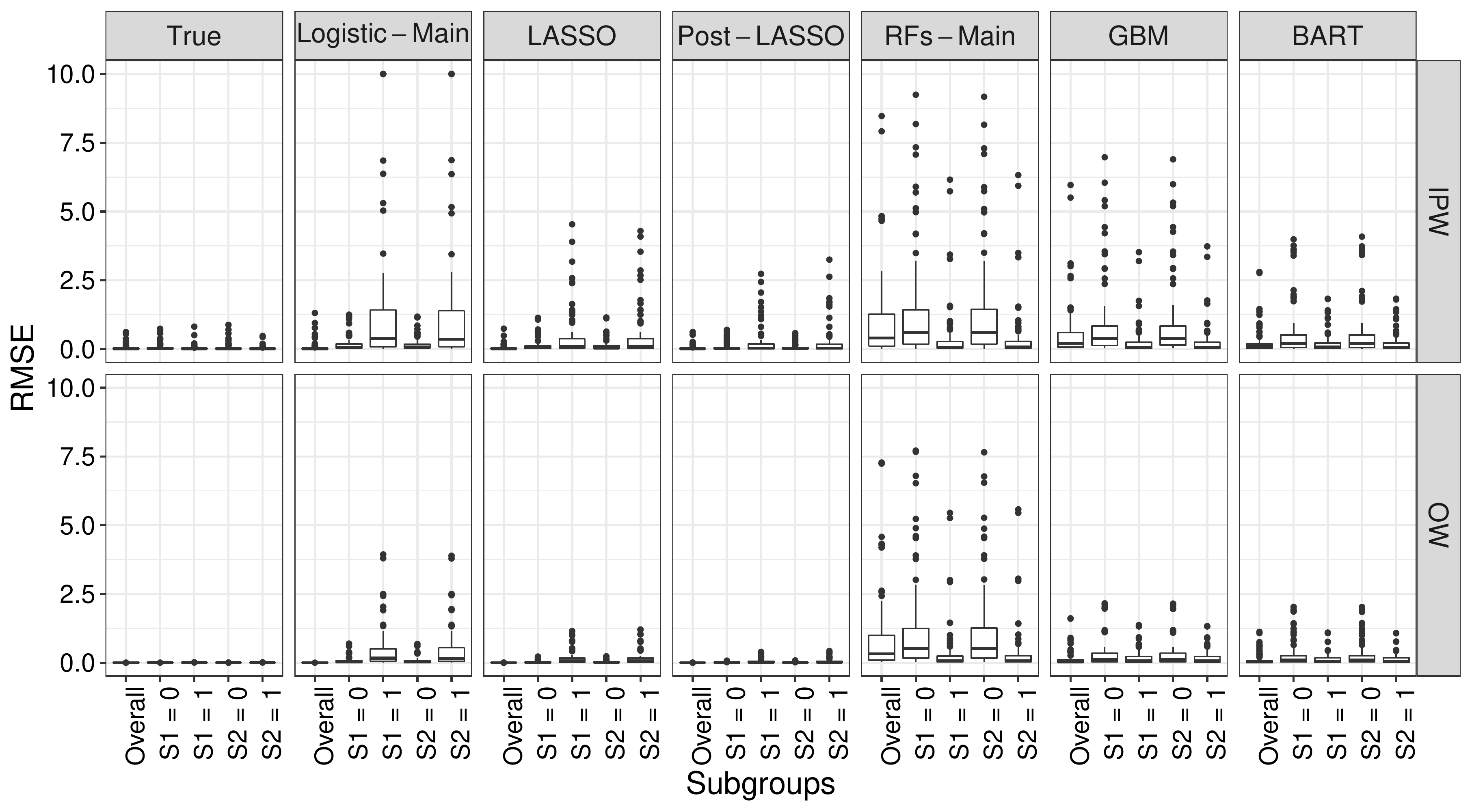}
    \caption{RMSE in estimating the overall WATE and the four subgroup S-WATE across different propensity models and weighting schemes. Values greater than 10 are truncated at 10. Each dot represents one of the 72 simulation scenarios.}
    \label{rmse}
\end{figure}

Covariate balance (AMSD), bias, and RMSE of the various estimators based on different postulated propensity score models and weighting schemes in the simulations are shown in Web Figure 2, Figure \ref{bias}, and Figure \ref{rmse}, respectively. 

%This section is distracting.  We can respond to
%reviewers if there is an issue.
%Note that the IPW and OW estimators target at %different estimands, and thus it is more meaningful %to compare bias or RMSE in relative rather than %absolute terms. Nonetheless, in our simulation %settings, the overall and all subgroup-specific true %ATO and ATE are very similar. Therefore, in %Figure~\ref{bias} and \ref{rmse}, we present the %bias and RMSE in absolute terms, which lead to %similar conclusion in terms of the relative %performance between different methods.

\emph{Balance.} From Web Figure 2, OW estimators achieve better covariate balance than IPW estimators across all propensity score models. The true propensity score model and OW achieves perfect balance for the true confounders in all subgroups. This is expected given OW's exact balance property for any included covariate-subgroup interactions (proposition \ref{prop:ps_balance}). Within the same weighting scheme, the LASSO and Post-LASSO model perform similarly, resulting in smaller ASMDs than the other methods. The Logistic-Main leads to satisfactory balance in the overall sample and the baseline subgroups (i.e. $S_1=0$ and $S_2=0$), but fails to balance the covariates in the $S_1=1$ and $S_2=1$ subgroups, particularly when paired with IPW. The RF models result in inferior balance performance (measured using ASMDs), occasionally leading to severe subgroup imbalances. BART and GBM perform similarly, which lie between the Logistic-Main and the LASSO models.

\emph{Bias.} From Figure~\ref{bias}, we can see that OW results in lower bias than IPW, for each propensity score modeling approach, both the overall and the subgroup effects. Between the different propensity score models, the pattern follows closely the degree of covariate imbalance. We find that OW-pLASSO returns the smallest bias within each subgroup and overall. LASSO is slightly inferior to Post-LASSO, likely due to the shrinkage induced bias. The common practice of using Logistic-Main IPW overestimates treatment effect in the baseline subgroups and greatly underestimates treatment effect in the $S_1=1$ and $S_2=1$ subgroups. If the same estimated propensity scores are paired with OW, the resulting estimates are much closer to the truth, and the bias for subgroups $S_1=1$ and $S_2=1$ is reduced to half. BART and GBM perform slightly better than the Logistic-Main and RFs model. Web Figure 3-4 provides more details of subgroup bias across a range of settings. Specifically, we find that the Logistic-Main IPW is much more sensitive to the simulation parameter specification compared to the OW-pLASSO. For example, it leads to substantial bias in estimating S-ATE under scenarios with more confounders and stronger confounding effects (i.e. larger $P$ and $\psi$, larger $\gamma$ and $\kappa$ values). %\noteG{I would make the following changes in the plot: (a) drop the color and the color legend -- you already have each method shown in a separate panel, (b) turn point plot to boxplots since it would show the range of each method more clearly (\texttt{geom\_boxplot} if you use ggplot), and (c) I would plot the absolute bias instead of the bias. Of course, feel free to ignore these recommendations. (a) and (b) also for RMSE plot}

\emph{RMSE.} From Figure~\ref{rmse} we can see that, with the same propensity score model, the RMSE is generally higher for IPW than for OW. This is expected, due to (i) the improved balance and (ii) the minimum variance property of OW. Neither the Logistic-Main nor the RF models capture the interactions in the true PS model and consequently result in large biases and variances of subgroup effects. This suggests that the RF models under our chosen hyperparameter settings are inadequate in learning the interactions (when given main effects only) or performing variable selection (when given the fully-expanded design matrix including subgroup interactions), leading to inaccurate and noisy treatment estimates. In contrast, LASSO coupled with OW provides low bias and high efficiency. Post-LASSO further improves upon LASSO across all the simulation settings we explored. Similarly to the previous observations, magnitude of the RMSE from BART and GBM is between that from the LASSO and Logistic-Main model. Web Figure 5-6 demonstrate the RMSE of OW-pLASSO is invariant to regression coefficients, while larger $P$ and $\psi$, larger $\gamma$ and $\kappa$ values greatly increase the RMSE of the IPW main effect model.

To summarize, OW estimators achieve better covariate balance, smaller relative bias and RMSE than IPW estimators across various propensity score models. 
The proposed method (OW-pLASSO) leads to low bias and high efficiency in estimating subgroup causal effects, suggesting LASSO successfully selects the important subgroup-covariate interactions across simulation scenarios. In contrast, the standard Logistic-Main as well as  alternative machine learning models for the propensity scores lead to large bias and RMSE in estimating the subgroup causal effects, particularly under moderate and strong confounding. %In addition, their performance is much more sensitive to specification of simulation hyperparameters.

\section{Application to COMPARE-UF}
\label{sec:application}
We now apply the proposed method to our motivating study of myomectomy versus hysterectomy in the 35 pre-specified subgroups of COMPARE-UF. In Figure \ref{fig:ConnectS}(c) the balance based on ASMD is substantially improved by OW-pLASSO, though still not perfect. This improvement in balance does not come at the expense of variance. Both overall and within subgroups, the variance inflation metric is lower with OW-pLASSO than the standard main effects logistic regression (panel (a)).  To save space in the comparison of methods we only show 6 subgroups.  Additional results for all subgroups were similar and are available in the Web Appendix 2.4. The only subgroup for which good balance was not achieved is age less than 35, though it was improved compared to the other methods. The challenge in balancing this subgroup is not surprising given the limited sample size and extreme imbalances that were initially present. We recommend that comparative statements about this subgroup are made very cautiously.

\begin{figure}[ht]
    \centering
    \includegraphics[width=7in]{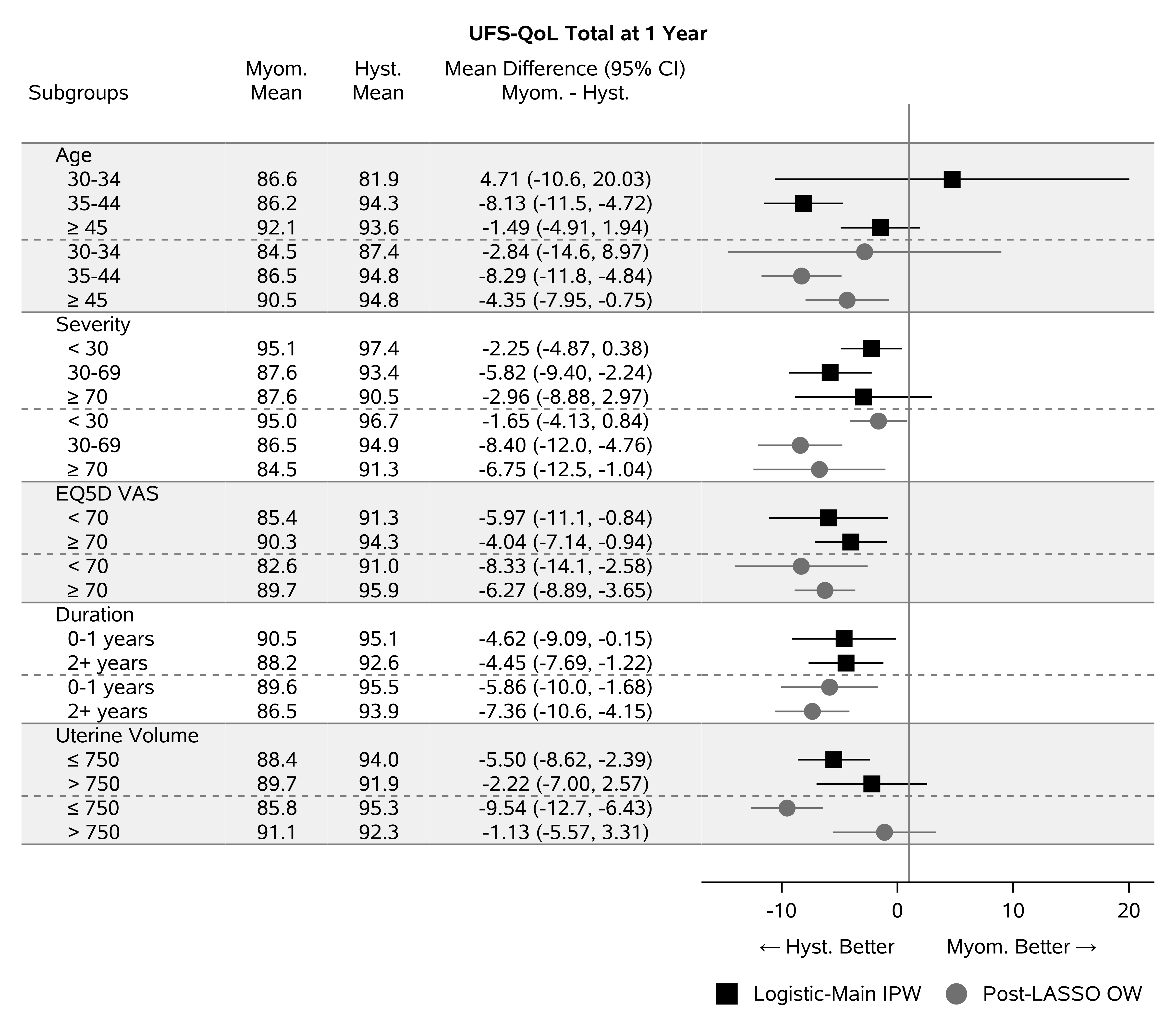}
    \caption{Estimates and 95$\%$ confidence intervals for treatment comparison of Myomectomy to Hysterectomy. Weighted means are reported and then contrasted.}
    \label{figow}
\end{figure}

Figure \ref{figow} displays estimated treatment effects for the primary quality of life endpoint, UFS-QOL score one year after the procedures with $95\%$ confidence intervals based on the robust sandwich variance estimator.\citep{austin2017performance} The proposed method, OW-pLASSO is compared to the standard Logistic-Main IPW. In some subgroups, including many of those not shown, the results of OW-pLASSO confirm those of Logistic-Main IPW.  However, some potentially important signals arise. OW-pLASSO reveals different treatment effects in the subgroups defined by baseline symptom severity. Individuals with mild symptom severity (<30) at baseline have similar outcomes with hysterectomy or myomectomy, whereas subgroups with higher initial symptoms (30-69, >70) receive a larger improvement in overall quality of life with hysterectomy. This is expected clinically, as hysterectomy entirely eliminates symptoms whereas symptoms can recur with myomectomy. Those with the greatest initial symptoms would have the most to gain.  The results of Logistic-Main IPW did not detect this difference. This is consistent with Figure \ref{fig:ConnectS} where covariate imbalances after weighting by Logistic-Main IPW were corrected by OW-pLASSO. A similar pattern was observed for the subgroups based on uterine volume. OW-pLASSO indicated that women with lower uterine volume had significantly larger benefits from hysterectomy. This result is not immediately intuitive, but may be related to the fact that women with lower uterine volume also had higher pain and self-consciousness score at baseline and therefore more to gain from a complete solution. This finding was obscured by Logistic-Main IPW because large imbalances in the baseline covariates favored myomectomy. 

\begin{figure}[ht]
    \centering
    \includegraphics[width=7in]{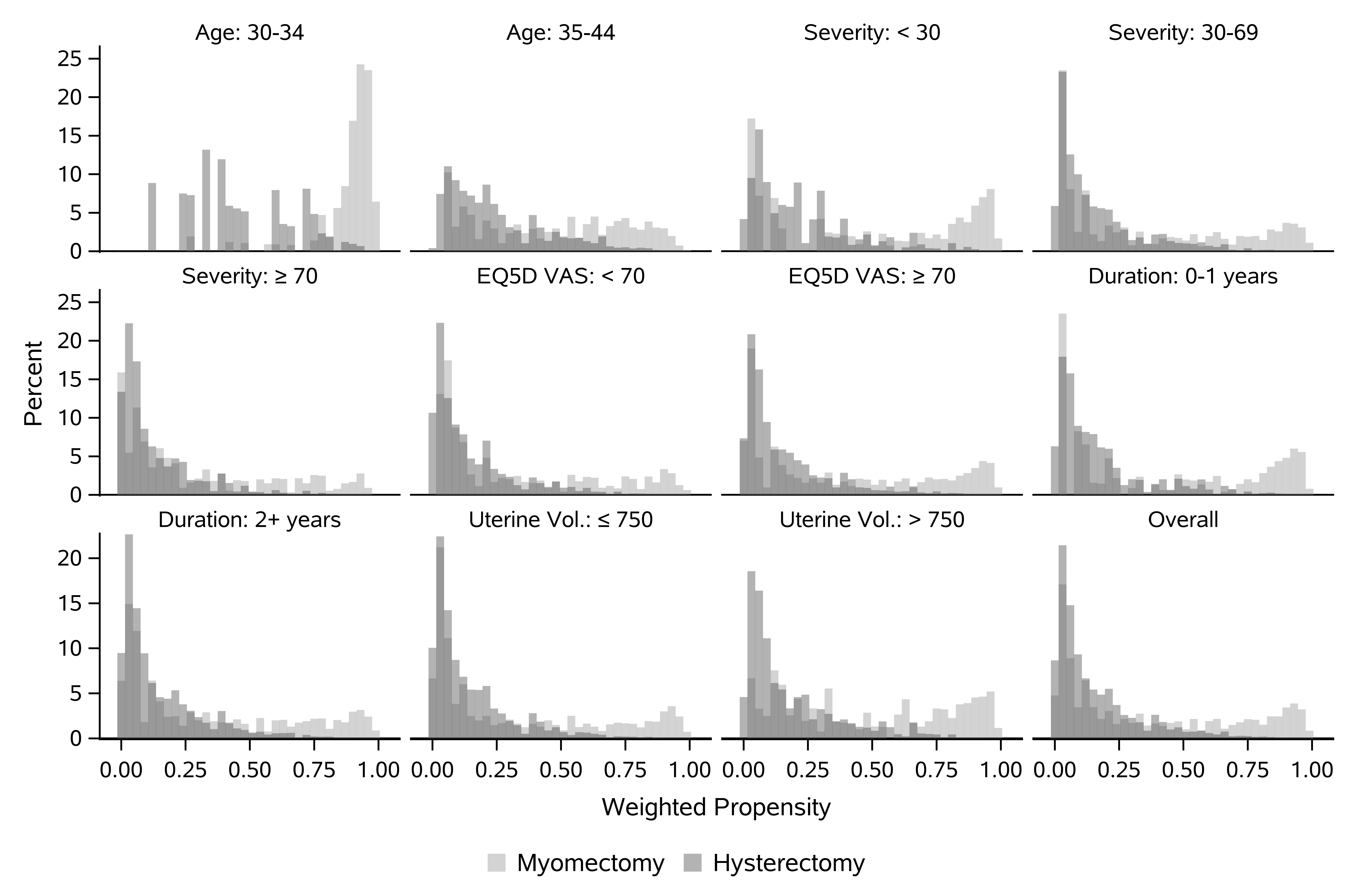}
    \caption{Propensity score distributions by treatment after weighting, by Logistic-Main IPW.}
    \label{fig:kde_ipw}
\end{figure}

\begin{figure}[ht]
    \centering
    \includegraphics[width=7in]{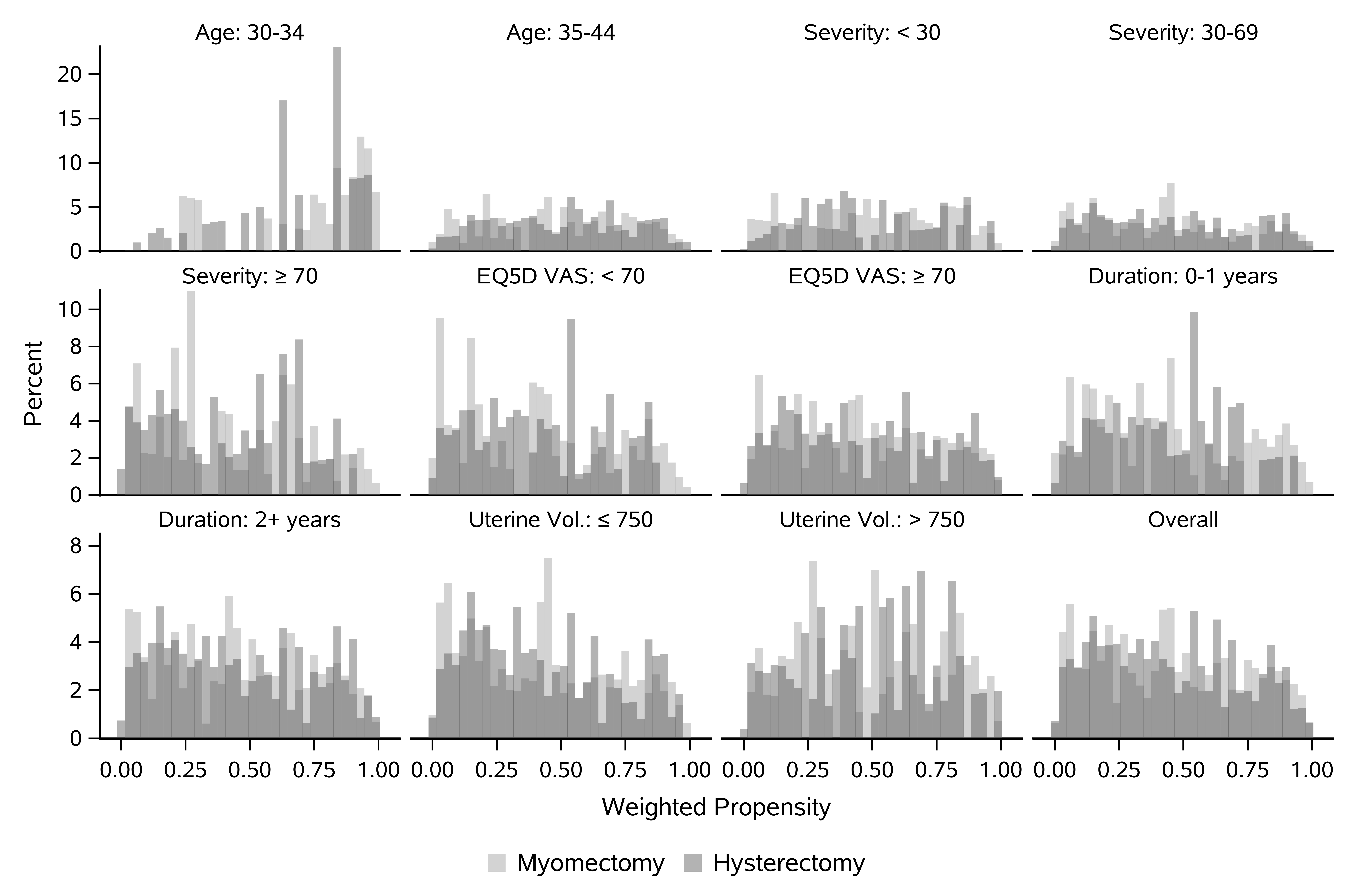}
    \caption{Propensity score distributions by treatment after weighting, by OW-pLASSO.}
    \label{fig:kde_ow}
\end{figure}

The COMPARE-UF data exemplify an additional advantage of OW-pLASSO, in the creation of a clinically relevant target population that emphasizes patients who are reasonably comparable, for all subgroups (S-ATO). To illustrate the shift in target population we display the propensity score distributions by subgroups after weighting. Figure \ref{fig:kde_ipw} illustrates two features of Logistic-Main IPW: (1) IPW has not made the hysterectomy and myomectomy groups similar; (2) The cohort is dominated by individuals at the extremes, with propensity values near 0 or 1. In contrast, the distributions in Figure \ref{fig:kde_ow} (resulting from OW-pLASSO) are mostly overlapping for hysterectomy versus myomectomy and emphasize people with propensity scores away from 0 and 1. While Logistic-Main IPW could be improved by iterative corrections, such as range trimming, or adapting the propensity score model, these steps would be cumbersome in COMPARE-UF to implement manually across 35 subgroups. Instead, OW-pLASSO automatically finds a population at clinical equipoise, for whom comparative data are most essential, across all subgroups. The resulting overlap cohort is displayed through a weighted baseline characteristics table in Web Appendix 2.4.  

\section{Discussion}
\label{sec:discussion}
As researchers look for real world evidence of comparative effectiveness in increasingly diverse and heterogeneous populations, it is crucial to advance appropriate methods for causal subgroup analysis with observational data.  In this paper we developed a suite of propensity score weighting methods and visualization tools for such a goal. We showed that it is essential to balance covariates within a subgroup, which bounds the estimation bias of subgroup causal effects. We further proposed a method that aims to balance the bias-variance trade-off in causal subgroup analysis. Our method combines Post-LASSO for selecting the propensity score model and overlap weighting for achieving exact balance within each subgroup and efficiency. We conducted extensive simulations to examine the operating characteristics of the proposed method. We found that pairing Post-LASSO with overlap weighting performed superior to several other commonly used methods in terms of balance, precision and stability. Our method provides one set of weights that can be used for both population average and subgroup-specific treatment effect estimation. The coupling of substantive knowledge about pre-specified subgroups, to generate candidate interactions, as well as machine learning for variable selection, may not only improve SGA but also the validity of the propensity score model for population average comparisons. As we move beyond SGA, using the knowledge of pre-specified subgroups to build the propensity score model may reduce bias in a range of propensity-score-based HTE methods. %\comment{Should we add some references here? }

We emphasized SGA with pre-specified subgroups in observational studies, while alternative methods and settings for HTE are rapidly developing. For example, Luedtke and van der Laan\cite{luedtke2017evaluating} showed that studying the additive treatment effect in SGA is similar to solving an optimization question when estimating the mean outcome. Recent research further recommends to select optimal subgroups based on the outcome mean difference between the effects and move away from one-covariate-at-a-time type of SGA.  \cite{vanderweele2019selecting} Similar to their idea, our method simultaneous uses all important covariates to make decisions.

The proposed methods maintain the causal inference principle of separating study design from analysis of outcomes. These methods allow an analyst to thoroughly investigate the model adequacy and balance without risk of being influenced by observing various treatment effects. Recent developments in causal inference are moving to incorporate information on the outcome in the propensity score estimation. \cite{shortreed2017outcome} When the candidate list of covariates is large, and investigators are not able to prioritize covariates, using the outcome data may be helpful. Future research could adapt the proposed method to incorporate outcome information.

We also designed a new diagnostic graph---the Connect-S plot---that allows visualizing subgroup balance for a large number of subgroups and covariates simultaneously. We hope the Connect-S plot and the associated programming code would facilitate more routine check of subgroup balance in comparative effectiveness research.    

The R and SAS code with implementation details used in this paper are provide at: \url{https://github.com/siyunyang/OW\_SGA}.

%\backmatter

\section*{Acknowledgments}
This research is supported by the Patient-Centered Outcomes Research Institute (PCORI) contract ME-2018C2-13289. The COMPARE-UF study, which constitutes our motivating example, was supported by the Agency for Healthcare Research and Quality grant RFA-HS-14-006. The contents of this article are solely the responsibility of the authors and do not necessarily represent the view of PCORI nor AHRQ.  We appreciate the clinical input and motivating questions from COMPARE-UF PI Evan Myers and COMPARE-UF investigators.  

% \subsection*{Financial disclosure}

%\subsection*{Author contributions}

%\subsection*{Conflict of interest}
%The authors declare no potential conflict of interests.

%\section*{Supporting information}
%The following supporting information is available as part of the online article:
%\noindent

%\appendix

%\nocite{*}% Show all bib entries - both cited and uncited; comment this line to view only cited bib entries;
\bibliography{SGAOW}%

% \clearpage

\end{document}